\documentclass[arxiv,superscriptaddress,twocolumn]{revtex4-1}

\usepackage{lineno,hyperref}

\usepackage{graphicx}
\usepackage{bmpsize}
\usepackage{amsfonts}
\usepackage[usenames, dvipsnames]{color}

\newcommand{\ben}{\begin{eqnarray}}
\newcommand{\een}{\end{eqnarray}}

\newcommand{\bef}{\begin{figure}[h!bt]\centering}
\newcommand{\eef}{\end{figure}}
\newcommand{\bet}{\begin{table}[hbt]\centering}
\newcommand{\eet}{\end{table}}

\begin{document}
\title{Magnetic order induces symmetry breaking in the single crystalline orthorhombic CuMnAs semimetal}
\author{Eve Emmanouilidou}
\affiliation{Department of Physics and Astronomy and California NanoSystems Institute, University of California, Los Angeles, CA 90095, USA}
\author{Huibo Cao}
\affiliation{Quantum Condensed Matter Division, Oak Ridge National Laboratory, Oak Ridge, TN 37831, USA}
\author{Peizhe Tang}
\affiliation{Department of Physics, McCullough Building, Stanford University, Stanford, California 94305-4045, USA}
\author{Xin Gui}
\affiliation{Department of Chemistry, Louisiana State University, Baton Rouge, Louisiana 70803, USA}
\author{Chaowei Hu}
\affiliation{Department of Physics and Astronomy and California NanoSystems Institute, University of California, Los Angeles, CA 90095, USA}
\author{Bing Shen}
\affiliation{Department of Physics and Astronomy and California NanoSystems Institute, University of California, Los Angeles, CA 90095, USA}
\author{Junyi Wu}
\affiliation{Department of Physics and Astronomy and California NanoSystems Institute, University of California, Los Angeles, CA 90095, USA}
\author{Shou-Cheng Zhang}
\affiliation{Department of Physics, McCullough Building, Stanford University, Stanford, California 94305-4045, USA}
\affiliation{Stanford Institute for Materials and Energy Sciences, SLAC National Accelerator Laboratory, Menlo Park, California 94025, USA.}
\author{Weiwei Xie}
\affiliation{Department of Chemistry, Louisiana State University, Baton Rouge, Louisiana 70803, USA}
\author{Ni Ni}
\email{Corresponding author: nini@physics.ucla.edu}
\affiliation {Department of Physics and Astronomy and California NanoSystems Institute, University of California, Los Angeles, CA 90095, USA}

\begin{abstract}
Recently, orthorhombic CuMnAs has been proposed to be a magnetic material where topological fermions exist around the Fermi level. Here we report the magnetic structure of the orthorhombic Cu$_{0.95}$MnAs and Cu$_{0.98}$Mn$_{0.96}$As single crystals. While Cu$_{0.95}$MnAs is a commensurate antiferromagnet (C-AFM) below 360 K with a propagation vector of \textbf{k} = 0, Cu$_{0.98}$Mn$_{0.96}$As undergoes a second-order paramagnetic to incommensurate antiferromagnetic (IC-AFM) phase transition at 320 K with \textbf{k} = (0.1,0,0), followed by a second-order IC-AFM to C-AFM phase transition at 230 K.
 In the C-AFM state, the Mn spins order parallel to the $b$-axis but antiparallel to their nearest-neighbors with the easy axis along the $b$ axis. This magnetic order breaks $R_y$ gliding and $S_{2z}$ rotational symmetries, the two crucial for symmetry analysis, resulting in finite band gaps at the crossing point and the disappearance of the massless topological fermions. However, the spin-polarized surface states and signature induced by non-trivial topology still can be observed in this system, which makes orthorhombic CuMnAs promising in antiferromagnetic spintronics.
\end{abstract}

\date{\today}
\maketitle

Dirac cones have been proposed and observed in many non-magnetic materials, including Cd$_3$As$_2$ \cite{Borisenko2014, Neupane2014} and Na$_3$Bi \cite{Liu2014, Xiong2015}.
By breaking inversion symmetry ($\mathcal{P}$) or time-reversal symmetry ($\mathcal{T}$), a Dirac point can be split into a pair of Weyl points. To break $\mathcal{T}$, we can either apply an external magnetic field or use the spontaneous magnetic moment inside the material. For the latter case, the correlation between spontaneous magnetism and Weyl fermions has been studied in the AMnPn$_2$ (A = rare earth or alkali earth and Pn = Sb or Bi) system  \cite{Wang2011, Wang2012, Wang2013, cava, Liu2015, Liu2016, Masuda2016, Huang2017} and the half-Heusler compound GdPtBi \cite{Hirschberger2016, Suzuki2016}. Recently, CuMnAs was proposed to be an interesting material with non-trivial topology. CuMnAs has two polymorphs; the tetragonal (TET) CuMnAs, which crystalizes in the space group $P4/nmm$, and the orthorhombic (ORT) CuMnAs crystalizing in the non-symmorphic $Pnma$ space group. The TET phase consists of alternating layers of edge-sharing CuAs$_4$ and MnAs$_4$ tetrahedra. It has been proposed to be a candidate with favourable applications in spintronics \cite{film1, film2} and a topological metal-insulator transition driven by the N\'{e}el vector \cite{prl}. On the other hand, the ORT phase consists of a 3D network of edge-sharing CuAs$_4$ and MnAs$_4$ tetrahedra (Fig. 2(c)), where the Mn atoms form a 3D distorted honeycomb lattice (Fig. 2(d)). ORT CuMnAs was proposed to be an antiferromagnetic topological semimetal when the spin-orbit coupling is fully considered \cite{shoucheng, prl}. In such a system, two gapless points, named as coupled Weyl fermions, are robust if the combination of $\mathcal{PT}$ is reserved and the non-symmorphic screw symmetry $\rm S_{2z}$ is not broken. Thus, the anti-ferromagnetic ORT CuMnAs provides an ideal system to study the interplay between antiferromagnetism (AFM) and Dirac fermions \cite{shoucheng}. In this paper, we will focus on the ORT CuMnAs. We experimentally determine its magnetic order, which breaks the $\mathcal{T}$ and $\mathcal{P}$ symmetries but keeps their combination $\mathcal{PT}$. We further show that this magnetic order will cause ORT CuMnAs to host interesting topological phase with spin-polarized surface states.

CuMnAs single crystals were grown via the high temperature solution method with Bi as the flux \cite{paul, supp}. The resistivities of the ORT single crystals are around tenths of m$\Omega$-cm and show metallic behavior. We observed two types of temperature-dependent resistivity behaviors. Figs. 1(a)-(b) show the normalized resistivity curves, $\rho(T)/\rho(400\rm K)$, representative of each type of behavior. Piece A (PA) shows a resistivity drop with a slope change at 360 K, suggesting the existence of one phase transition. The derivative of resistivity, d$\rho/\rm{d}T$, shows a sharp kink at 360 K. On the other hand, piece B (PB) shows two resistive anomalies, suggesting the occurrence of two successive phase transitions. The d$\rho/\rm{d}T$ plot indicates that one kink appears around 320 K and the other occurs around 230 K. Table SI includes a summary of the relation between the number of resistive anomalies and the growth condition. The inset of Fig. 1(a) shows the field dependent Hall resistivity $\rho_{yx}$(H) of PA at 2 K and 100 K. $\rho_{yx}$ is positive, indicating that holes dominate the transport. It is linearly proportional to H and shows almost no temperature dependence, suggesting the validity of the single band model here. Based on $n= $B/$e\rho_{yx}$, the estimated carrier density is $\approx $6.5$\times 10^{20}/$cm$^3$. This value is significantly greater than Dirac semimetals Cd$_3$As$_2$ \cite{Liang2015}, Na$_3$Bi \cite{Xiong2016} and Weyl semimetal TaAs \cite{taas}, but comparable to the Dirac nodal-line semimetal candidates ZrSiSe \cite{zrsise} and CaAgAs\cite{Eve2017}.
\begin{figure}
  \includegraphics[width=3.4in]{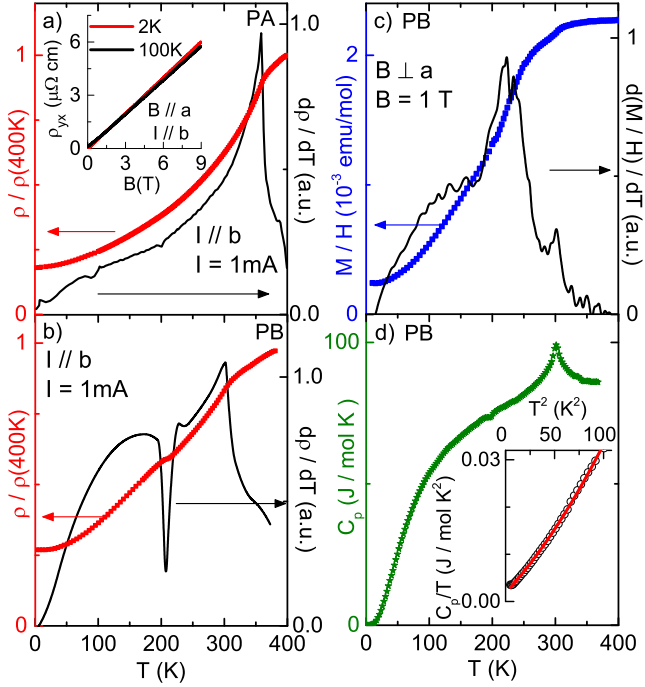}
  \caption{PA: (a) The normalized resistivity $\rho$/$\rho$(400 K) and its derivative d$\rho$/d$T$ vs. T. Inset: The Hall resistivity $\rho_{yx}$ vs. T. PB: (b)-(d): (b): The normalized resistivity $\rho/\rho$(400 K) and d$\rho$/d$T$ vs. T. (c): The susceptibility M/H and d($M/H$)/d$T$ vs. T. (d) The heat capacity C$_p$ vs. T. Inset: C$_p$/T vs. T$^2$.}
  \label{fig:}
\end{figure}

As a representative, the temperature dependent susceptibility ($M/H(T)$) and heat capacity ($C_p(T)$) of PB are presented in Figs. 1(c)-(d). Two slope changes can also be observed in the $M/H(T)$ data, which can be clearly seen in d($M/H$)/d$T$. From 300 K to 400 K, the highest temperature we measured, the $M/H(T)$ data is almost temperature independent, showing no Curie-Weiss behavior. No metamagnetic phase transition is observed with H // $b$ and $c$ directions up to 7 T. The $C_p(T)$ data show only one heat capacity jump around 320 K without any anomaly at 230 K, suggesting that the phase transition at 230 K is most likely a transition between two ordered phases. Since both phase transitions are at high temperatures, we fitted the $C_p/T$ data from 2 K to 10 K using $C_p = \gamma T  + \alpha T^3 + \beta T^5$, where the first term refers to the electronic heat capacity and the rest to the low temperature lattice heat capacity. We deduced a Sommerfeld coefficient $\gamma$=1.88 mJ mol$^{-1}$ K$^{-2}$ which indicates a small density of states at the Fermi level for the ORT CuMnAs.

To shed light on the difference in physical properties between PA and PB, single crystal X-ray and neutron diffraction measurements were performed on these two pieces to investigate their structural properties. No structural phase transition is detected down to 100 K. Combined with the SEM-EDX data, which gives Cu$_{0.98(3)}$Mn$_{0.98(4)}$As$_{1.02(4)}$ for both PA and PB, by examining the refinements based on X-ray and neutron data via various models, we are convinced that both site vacancies and site disorders exist \cite {supp}. Tables SII and SIII summarize the refined crystal structure, atomic positions and site occupancies of PA and PB. The major difference between them is the stoichiometry. PA has fully occupied Mn sites with 5.0(2)\% of Cu site vacancies, leading to a chemical formula of Cu$_{0.95}$MnAs, while PB has vacancies in both Cu and Mn site with a chemical formula of Cu$_{0.98}$Mn$_{0.96}$As. Furthermore, revealed by neutron scattering, $\sim$5\% and 6\% of Cu and Mn site mixing also exist in PA and PB, respectively \cite{supp}. In the rest of the paper, we will denote Cu$_{0.95}$MnAs as PA and Cu$_{0.98}$Mn$_{0.96}$As as PB. The difference in the physical properties between PA and PB likely arises from the stoichiometry of the Mn and Cu sites.

\begin{figure}
 \includegraphics[width=3.4in]{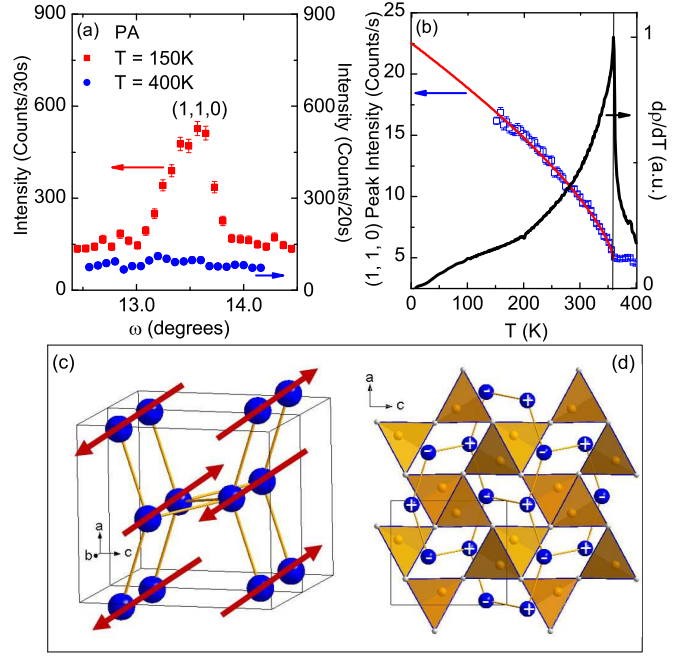}
 \caption{(a) The (1,1,0) intensity vs. $\omega$ for PA. (b) A comparison between the (1,1,0) peak intensity and the d$\rho$/d$T$ vs. $T$. The red line is the power law fit, see text. (c) The magnetic structure of PA in the C-AFM state. Only the Mn sublattice is shown. (d) The view of the magnetic structure from the $b$ direction. Mn atoms are shown in blue. ``+" denotes spins pointing out of plane while ``-" denotes spin pointing in plane.}
 \label{fig:}
\end{figure}

\begin{figure}
 \includegraphics[width=3.4in]{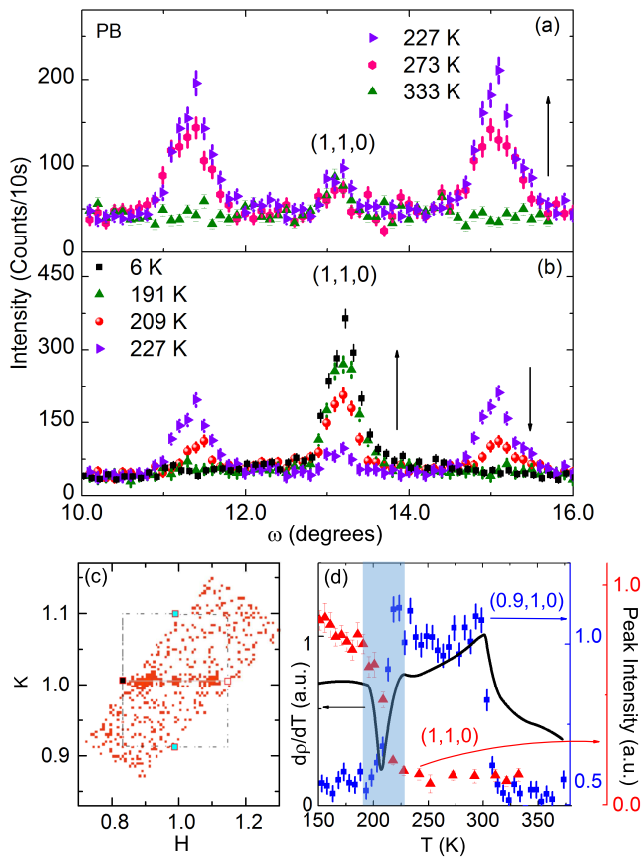}
 \caption{(a)-(b): The intensity vs. $\omega$ for PB. (c): A cut of the neutron scattering in the hk0 plane. (d) A comparison between the (0.9,1,0) peak intensity and d$\rho$/d$T$ vs. $T$. The light blue box marks the temperature region where C-AFM and IC-AFM compete and coexist.}
 \label{fig:}
\end{figure}

To unravel the nature of these phase transitions, single crystal neutron diffraction was performed using HB-3A single crystal neutron diffractometer at High Flux Isotope Reactor (HFIR) at Oak Ridge National Laboratory (ORNL). Figure 2 (a) presents the rocking curve scan at (1,1,0) on PA. The peak shown at (1,1,0) at 150 K indicates a long range commensurate AFM (C-AFM) order. The temperature dependent (1,1,0) peak intensity agrees well with the d$\rho$/d$T$, shown in Fig. 2(b). It suggests a second order AFM phase transition and can be fitted using the power law $I(T)/I_0=(M(T)/M_0)^2=A+(1-T/T_N)^{2\beta}$, where M$_0$ is the saturation moment. With $T_N=360$ K, the critical exponent is $\beta=0.35(3)$, which agrees with the $\phi^4$ model in 3D \cite{baker} and suggests the breakdown of the mean field theory ($\beta=0.5$) and thus a strong spin fluctuation near $T_N$. We refined the magnetic and nuclear structure of Cu$_{0.95}$MnAs together based on 76 effective magnetic reflections. $Pn^\prime ma$ is the only magnetic symmetry which can fit the data. The R-factor is 0.0508 and the goodness of fit is 6.08. Figures 2(c)-(d) show the refined C-AFM structure. Mn spins sit on a distorted honeycomb sublattice and order parallel to each other along the $b$ axis (Fig. 2 (c)) with the nearest-neighboring spins antiferromagnetically aligned to each other (Fig. 2(d)). This magnetic structure is the same as the one proposed theoretically in Ref. \cite{shoucheng}, but with the spin orientation along the $b$ axis. The refined magnetic moment at 150 K is 4.0(1) $\mu_B$/Mn.

\begin{figure}
 \includegraphics[width=3.2in]{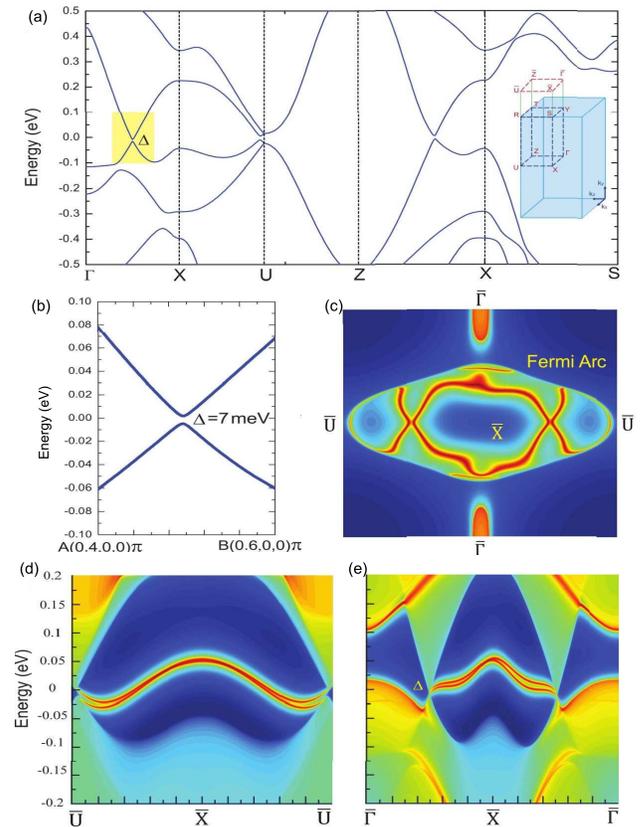}

 \caption{(a) The calculated band structures of the ORT CuMnAs with SOC and the magnetic structure shown in Fig. 2(c). Inset is the Brillouin zone and its projection to the (010) surface. $\Delta$ is the band gap of massive topological fermion along $\Gamma$X line. (b) The zoom-in view of the band structure marked by yellow box in (a). (c) The Fermi surface contour on the (010) surface at the calculated Fermi level. The corresponding electronic spectra along (d) $\bar{k_{x}}$=$\pi/a$ and (e) $\bar{k_{z}}$=0. The Fermi level is set to zero.}
 \label{fig:DFT}
\end{figure}
Figures 3(a) and (b) show the rocking curve scan at (1,1,0) on PB. The (1,1,0) is not allowed by the crystal structure symmetry in the ORT CuMnAs phase, and the non-zero intensity we observed above 320 K matches with 1.4\% half-lambda contamination at HB-3A (without the filter). In Fig. 3(a), magnetic satellite peaks begin to appear near the (1,1,0) as the sample is cooled below 320 K, suggesting incommensurate antiferromagnetism (IC-AFM). Figure 3(c) shows a cut in the hk0 plane at 227 K. We can clearly detect the three peaks shown in Fig. 3(a). The concentration of points at (0.9,1,0) and (1.1,1,0) indicates the presence of the incommensurate magnetic peaks, suggesting an incommensurate propagation vector \textbf{k} = (0.1,0,0). Upon further cooling of the sample below 230 K, we observe that both (0.9,1,0) and (1.1,1,0) peak intensities decrease while the (1,1,0) peak intensity starts to increase, indicating a competition between the C-AFM with the propagation vector \textbf{k} = 0 and IC-AFM. Below 190 K, both (0.9,1,0) and (1.1,1,0) peaks diminish whereas the (1,1,0) peak keeps increasing, suggesting the disappearance of IC-AFM. To better visualize the competition and coexistence, Fig. 3(d) shows the (0.9,1,0) and (1,1,0) peak intensities and d$\rho$/d$T$ as a function of temperature. We can see that Cu$_{0.98}$Mn$_{0.96}$As undergoes a second-order PM to IC-AFM phase transition at 320 K as well as a second-order IC-AFM to C-AFM phase transition at 230 K. IC-AFM competes and coexists with the C-AFM phase between 230 to 190 K and disappears below 190 K.
Based on 102 effective magnetic peaks, the refined C-AFM structure is the same as the one in Cu$_{0.95}$MnAs (Figs. 2 (c)-(d)) with the refined magnetic moment at 6 K as 4.3(2) $\mu_B$/Mn and a R-factor of 0.0555. The moment is smaller than 5 $\mu_B$/Mn, the theoretical saturation moment for Mn$^{2+}$.

In order to explore the electronic and topological properties of ORT CuMnAs with the magnetic orientations along $b$ axis, we calculate its bulk band structures and the corresponding (010) surface states, as shown in Fig. \ref{fig:DFT}. Due to the presence of the $\mathcal{PT}$ symmetry in the experimental C-AFM phase, every bulk state is double degenerate. Furthermore, the band inversion still exists in this system, thus the non-trivial topological properties can appear. Because the C-AFM order breaks the non-symmorphic gliding symmetry R$_y$ and screw symmetry $S_{2z}$, in contrast to the case with spin orientation along the $c$ axis \cite{shoucheng}, now the gapless coupled Weyl fermions disappear and the Dirac nodal line is fully gapped everywhere by SOC in the bulk Brillouin zone (BZ), as shown in Fig. \ref{fig:DFT}(a). But the gap size is quite small, for example, the band gap induced by SOC along $\Gamma$X line is just 7 meV, as indicated in Fig. \ref{fig:DFT}(b). Figure \ref{fig:DFT}(c) shows the spin-polarized surface states emerging from the gapped bulk states (see Fig. \ref{fig:DFT}(d) and (e)) on the (010) side surface. Due to the absence of rotation symmetries on the (010) surface, the Fermi surface contour at the Fermi level is asymmetric, and the spin-polarized surface states are gapped. This distinguishing character is different from surface states in topological insulators and Dirac semimetals. On the other hand, because the bulk Dirac fermions in this case are massive and the time reversal symmetry is broken, the fluctuations could resemble the dynamical axion field, which gives rise to exotic modulation of the electromagnetic field showing the similar signature of axion insulators \cite{shoucheng2}.

In conclusion, the Dirac antiferromagnetic semimetal candidates, ORT Cu$_{0.95}$MnAs and Cu$_{0.98}$Mn$_{0.96}$As single crystals, show no structural phase transitions down to 100 K. The magnetism is very sensitive to the stoichiometry of the Cu and Mn sites. Cu$_{0.95}$MnAs has a PM to C-AFM phase transition at 360 K while an intermediate IC-AFM state between PM and C-AFM states appears in Cu$_{0.98}$Mn$_{0.96}$As. In both C-AFM state, spins order parallel to one another along the $b$ axis, but antiparallel to their Mn nearest-neighbors with the moment around 4.3 $\mu_B$/Mn. The spin orientations are along the $b$ axis, which break both $S_{2z}$ and $R_y$ symmetries in the C-AFM state and gap the coupled Weyl nodes and Dirac nodal line, resulting in similar electromagnetic response to that of axion insulators. Furthermore, the presence of spin-polarized surface states on this AFM semimetal makes ORT CuMnAs to be a good candidate for the antiferromagnetic spintronic applications.

\vspace{2mm}

Work at UCLA was supported by the U.S. Department of Energy (DOE), Office of Science, Office of Basic Energy Sciences under Award Number DE-SC0011978. Work at ORNL$^\prime$s High Flux Isotope Reactor was sponsored by the Scientific User Facilities Division, Office of Basic Energy Sciences, DOE. The research at LSU was supported by the LSU-startup funding and Louisiana Board of Regents Research Competitiveness Subprogram under the Contract Number LEQSF (2017-20)-RD-A-08. PZ and SCZ acknowledge FAME, one of six centers of STARnet, a Semiconductor Research Corporation program sponsored by MARCO and DARPA. NN thanks Dr. Yilin Wang for useful discussion.


\begin{thebibliography}{99}
\bibitem{Borisenko2014} S. Borisenko, Q. Gibson, D. Evtushinsky, V. Zabolotnyy, B. Buchner, and R. J. Cava, Phys. Rev. Lett. \textbf{113}, 027603 (2014)
\bibitem{Neupane2014} M. Neupane, S.-Y. Xu, R. Sankar, N. Alidoust, G. Bian, C. Liu, I. Belopolski, T.-R. Chang, H.-T. Jeng, H. Lin et al., Nat. Commun. \textbf{5}, 3786 (2014)
\bibitem{Liu2014} Z. K. Liu, B. Zhou, Y. Zhang, Z. J. Wang, H. M. Weng, D. Prabhakaran, S.-K. Mo, Z. X. Shen, Z. Fang, X. Dai et al., Science \textbf{343}, 864 (2014)
\bibitem{Xiong2015} J. Xiong, S. K. Kushwaha, T. Liang, J. W. Krizan, M. Hirschberger, W. Wang, R. J. Cava, and N. P. Ong, Science \textbf{350}, 413 (2015)


\bibitem{Wang2011} K. Wang, D. Graf, H. Lei, S.W. Tozer, and C. Petrovic, Phys. Rev. B \textbf{84}, 220401 (2011)
\bibitem{Wang2012} K. Wang, D. Graf, L. Wang, H. Lei, S.W. Tozer, and C. Petrovic, Phys. Rev. B \textbf{85}, 41101 (2012)
\bibitem{Wang2013}  K. Wang, D. Graf, and C. Petrovic, Phys. Rev. B \textbf{87}, 235101 (2013)
\bibitem{cava} S. Borisenko, D. Evtushinsky, Q. Gibson, A. Yaresko, T. Kim, M.N. Ali, B. B\"{u}chner, M. Hoesch, and R.J. Cava, arXiv:1507.04847v2, Unpublished, (2015)

\bibitem{Liu2015} J.Y. Liu, J. Hu, Q. Zhang, D. Graf, H.B. Cao, S.M.A. Radmanesh, D.J. Adams, Y.L. Zhu, G. F. Cheng, X. Liu, et. al, arXiv:1507.07978v2, Unpublished, (2015)
\bibitem{Liu2016} J. Liu, J. Hu, H. Cao, Y. Zhu, A. Chuang, D. Graf, D.J. Adams, S.M.A. Radmanesh, L. Spinu, I. Chiorescu, and Z. Mao, Sci. Rep. \textbf{6}, 30525 (2016)
\bibitem{Masuda2016} H. Masuda, H. Sakai, M. Tokunaga, Y. Yamasaki, A. Miyake, J. Shiogai, S. Nakamura, S. Awaji, A. Tsukazaki, H. Nakao, et. al, Sci. Adv. \textbf{2}, (2016)
    \bibitem{Huang2017} S. Huang, J. Kim, W.A. Shelton, E.W. Plummer, and R. Jin, Proc. Natl. Acad. Sci. \textbf{114}, 6256 (2017)
\bibitem{Hirschberger2016} M. Hirschberger, S. Kushwaha, Z. Wang, Q. Gibson, S. Liang, C.A. Belvin, B.A. Bernevig, R.J. Cava, and N.P. Ong, Nat. Mater. \textbf{15}, 1161 (2016)
\bibitem{Suzuki2016} T. Suzuki, R. Chisnell, A. Devarakonda, Y.-T. Liu, W. Feng, D. Xiao, J.W. Lynn, and J.G. Checkelsky, Nat. Phys. \textbf{12}, 1119 (2016)

\bibitem{shoucheng} P. Tang, Q. Zhou, G. Xu, S.-C. Zhang, Nat. Phys. \textbf{12}, 1100 (2016)

\bibitem{MacA2012} F. MacA, J. Masek, O. Stelmakhovych, X. Marti, H. Reichlova, K. Uhlirova, P. Beran, P. Wadley, V. Novak, and T. Jungwirth, J. Magn. Magn. Mater. \textbf{324}, 1606 (2012)

\bibitem{prl} L. \u{S}mejkal, J. \u{Z}elezn\'{y}, J. Sinova, and T. Jungwirth, Phy. Rev. Lett. 118, 106402 (2017)
\bibitem{film1}  P. Wadley, V. Hills, M. R. Shahedkhah, K. W. Edmonds, R. P. Campion, V. Nov\"{a}k, B. Ouladdiaf, D. Khalyavin, S. Langridge, V. Saidl, et. al, Sci. Rep. \textbf{5}, 17079 (2015)
\bibitem{film2} P. Wadley, B. Howells, J. \u{Z}elezn\'{y}, C. Andrews, V. Hills, R. P. Campion, V. Novák, K. Olejn\'{i}k, F. Maccherozzi, S. S. Dhesi,et. al, science, 351, 587-590 (2016)
     \bibitem{paul}P. C. Canfield and Z. Fisk, Philos. Mag. B 65, 1117 (1992)
    \bibitem{supp} See supplementary materials.
\bibitem{Liang2015} T. Liang, Q. Gibson, M. N. Ali, M. Liu, R. J. Cava, and N. P. Ong, Nat. Mater. \textbf{14}, 280 (2014)
\bibitem{Xiong2016} J. Xiong, S. Kushwaha, J. Krizan, T. Liang, R. J. Cava, and N. P. Ong, Europhys. Lett. \textbf{114}, 27002 (2016)
\bibitem{taas} C. L. Zhang, Su-Yang Xu, Ilya Belopolski, Zhujun Yuan, Ziquan Lin, Bingbing Tong, Guang Bian, Nasser Alidoust, Chi-Cheng Lee, Shin-Ming Huang, et. al, Nat. Comm. 7, 10735 (2016)
\bibitem{zrsise} Jin Hu, Zhijie Tang, Jinyu Liu, Xue Liu, Yanglin Zhu, David Graf, Kevin Myhro, Son Tran, Chun Ning Lau, Jiang Wei, et. al, Phy. Rev. lett. 117, 016602 (2016)
\bibitem{Eve2017} E. Emmanouilidou, B. Shen, X. Deng, T.-R. Chang, A. Shi, G. Kotliar, S.-Y. Xu, and N. Ni, Phys. Rev. B \textbf{95}, 245113 (2017)
\bibitem{baker} G.A. Baker, B.G. Nickel, and D.I. Meiron, Phys. Rev. B \textbf{17}, 1365 (1978)
\bibitem{shoucheng2} Rundong Li, Jing Wang, Xiao-Liang Qi and Shou-Cheng Zhang, Nat. Phys. 6, 284 (2010)




\end{thebibliography}
\end{document}